# Modeling Hypermedia-Based Communication

Kryssanov[1*], V.V., Kakusho[2], K., Kuleshov[3], E.L., and Minoh[2], M.

[1] Information and Communication Sciences Dept., Ritsumeikan University, Kusatsu, Japan

[2] Academic Center for Computing and Media Studies, Kyoto University, Kyoto, Japan

[3] Computer Systems Dept., the Far-Eastern National University, Vladivostok, Russia

**Abstract**

In this article, we explore two approaches to modeling hypermedia-based communication. It is argued that the classical conveyor-tube framework is not applicable to the case of computer- and Internet- mediated communication. We then present a simple but very general system-theoretic model of the communication process, propose its mathematical interpretation, and derive several formulas, which qualitatively and quantitatively accord with data obtained on-line. The devised theoretical results generalize and correct the Zipf-Mandelbrot law and can be used in information system design. At the paper's end, we give some conclusions and draw implications for future work.

---

[*] The corresponding author; related correspondence should be sent to Kryssanov V.V., College of Information Science and Engineering, Ritsumeikan University, Kusatsu, Japan. Tel./fax: +81 (0)77-561-5936, e-mail: kvvictor@is.ritsumei.ac.jp, kryssanov@mm.media.kyoto-u.ac.jp

# 1. Introduction

Hypermedia is an approach to the development of information systems (see reference [4] for an overview). A hypermedia system stores and manages its information as a collection of nodes, (hyper)links, and scripts. Nodes are multiple representations of different media types – chunks of text, photographs, pictures, sounds, movies, etc. Links implement transitions (by association or connection) between nodes, while scripts, which are generalizations of links, are used to combine and control the diverse nodes in a digital document. Hypermedia-based communication may be thought of as the cognitive processing by a hypermedia system's user that includes selecting, organizing, and integrating information represented over time in the networks of nodes.

Hypermedia-based communication differs from "natural" word-based communication in terms of both structure and dynamics. The latter process is usually defined as a linear, single-channel/medium, receiver-focused, but source-driven transfer of information. In contrast, the former is fragmented, multi-channel/media, network-focused and receiver-driven. Owing to the diversity of deployed representations, one might also claim that the former is a generalization of the latter. It is remarkable that, while these two are the main types of communication on the present Internet, only word-based communication has been receiving somewhat adequate technological support, since the existing interface and networked systems rest on the "conveyor-tube" information-theoretic model and its mathematical interpretations developed in quantitative linguistics half a century ago [21, 23, 14].

The conveyor-tube model deals with the efficiency of transmitting, as well as coding and decoding of the information communicated from an active source/sender to the target/receiver, and it builds on several assumptions about properties of the communication process. Most interesting for us are the assumptions about the statistical homogeneity of communication (i.e. about a linear discourse of a single speaker or writer) and about the negligences of semantic effects (including those due to context) and redundancy (i.e. before transmission, information is to be efficiently encoded) in communication. The model offers an explanation to several important probabilistic phenomena, including Zipf's "second law" (also known as the discrete Pareto distribution; see extensive bibliography in reference [8]) establishing a relation between frequency and count of words in a word-by-word communication, that are currently employed by search engines, automatic text analysis and indexing tools, and other applications sustaining communication on the Internet. This explanation is, however obviously, not valid or relevant for the case of hypermedia-based communication with its non-linear fragmented discourse and manifold information source-nodes, in effect semantic interdependence-links, and inefficient encoding of information (e.g. due to an overlap of different media types).

Unfortunately, numerous papers on computer- and Internet- mediated communication do little to fill this gap in theoretical understanding and modeling capability, whereas computational models of linguistics, which are routinely used in computer science, are rarely compared with and do not generally match the Internet-specific data available on-line. As a result, designing hypermedia systems currently appears an art rather than an engineering process, and studies of hypermedia-based communication remain phenomenological and atheoretical: even though a great deal is known about a number of isolated phenomena, such as characteristics of nodes, links, and entire digital documents, there is a poor understanding of how these phenomena interact and influence communication within a wider frame [18].

In this article, we are concerned with four points:

1. The absence of a credible theoretical framework is a design fundamental problem of all hypermedia applications.

2. Communication should generally be studied as a social phenomenon rather than individual one-to-one interactions.

3. Complex system theory provides us with a way to model hypermedia-based communication with a precision sufficient for many analytic and practical purposes.

4. Empirical validation of communication models is unavoidable, and it cannot be replaced by theoretical speculations, no matter how plausible or logical they appear.

In the following section, we describe a conceptual framework for modeling hypermedia-based communication. The framework takes a system-theoretic perspective and defines communication as a mutually-orienting perturbation of coupled autonomous systems. Section 3 gives a mathematical interpretation of this definition: we derive, with a bare minimum of assumptions, a statistical model that explains fluctuations of frequencies of representations, such as words or links, in a (digital) document. In Section 4, we test the proposed model against real data and show that it gives a statistically sound tool for the analysis of information systems and for the studies of communication as well, which is far superior (in the statistical sense) to the classical model. Finally, Section 5 discusses the study findings, and Section 6 concludes the paper.

At this point, we would like to mention one caution. The goal of the presented study is not a reinterpretation of the Zipf-Mandelbrot idea of the "least effort" or efficient information coding [14]. Nor do we aim to improve Simon's (power law) generative model of word frequencies [23]. Instead, the paper comes from an entirely different explanation of the communication process. Our chief focus is, therefore, by no means a critique itself or advancement of the classical theories of word frequencies and communication dynamics. The reader interested in the latter topic is directed to reference [8, 16, 11].

## 2. Understanding the Communication Process

Most contemporary models of the communication process utilize the old *Stimulus-Response Model* of behavioral psychology. There is a *sender* who encodes her or his perception or idea into a message. There is also a *receiver* who decodes the message and provides feedback. Communication is portrayed as the transmission of meaning – the perception or idea – from the sender to the receiver.

The conveyor-tube model gives a technological interpretation of the Stimulus-Response Model [21]. It shows the role of the *medium*/channel (e.g. a language) in which the idea or perception is packaged as the cause of possible alterations of the information transmitted. It also introduces a third party – an observer, who intervenes in the process by determining the successfulness of communication through comparison of original and received meanings.

The strongest point of the conveyor-tube framework is that it allows for statistically modeling the communication process, providing one can measure "the amount of" meaning, the efficiency of its coding, or determine the "noise properties" (i.e. to what extent it would alter the conveyed perception or idea) of the medium. The specificity of hypermedia-based communication is, however, such that at every moment, there are numerous potential senders and several media, and the receiver's meaning is a result of the cognitive processing of a number of miscellaneous and often incomplete messages, rather than a uniform decoding process. The principal problem is thus that the conveyor-tube model does not account for the fact that meaning is not transmitted from a single source, but created in the mind of the receiver, based at best in part on the sender's message but also – on social and idiosyncratic parameters of the communication situation, and yet sometimes regardless of the sender's original intention.

In an attempt to compensate for this flaw, the Stimulus-Response Model has been generalized to the case of non-human communication by reviewing the process of communication from positions of evolutionary structuralism [15]. Communicative activities constitute a class of observed behavior of self-organizing systems. The principal property of a self-organizing system (e.g. human, other animal, or some machine) is its autonomy in respect to the environment: the inner state of such a system at any time is determined solely by its structure and previous state. Environmental perturbations can only be a potential cause for the changing of the system state, and the system cannot be controlled from the outside. Hence, all observed behavior – the output – of a self-organizing system is a result of its inner state and history. Through behavior, the system can interact with the environment that may cause it to change its structure, so that the system becomes structurally coupled with the environment. It is said that the coupled system undergoes self-adaptation, when the system and its dynamic environment mutually trigger their inner states. The self-adaptation processes of several systems embedded in the same environment may become coupled, recursively acting through their own states. All the possible changes of states of such systems, which do not terminate this coupling, establish a consensual domain. Behavior in a consensual domain is mutually orienting. Perhaps most fundamentally [17]:

*Communication is the (observed) behavioral coordination developed from the interactions between autonomous (self-organizing) systems in the consensual domain.*

Putting it formally, communication as the mutually orienting interaction process can be represented in the form of an *n*-tuple of pairs of simultaneous equations [10]:

$$\begin{cases} R_{k,j+1} = \mathbf{E}_k(S_{k,j}, t), \\ S_{k,j+1} = \mathbf{I}_k(R_{k-1+n\delta_{in}, j+1}, t), \end{cases} \quad (1)$$

where $R_{k,j}$ denotes a (local) realization of $S$ a (macro) state of a $k$-type system at $[t_{j-1}, t_j]$ a discrete time interval, $t_{j-1} < t_j, j=1,2,\ldots$; $k=n-i+1$, $i=1,\ldots,n$; $n$ determines the depth of coupling, $\delta_{in}$ is the Kronecker delta: $\delta_{in}=1$ iff $i=n$, and $\delta_{in}=0$, otherwise; $\mathbf{E}$ and $\mathbf{I}$ are time-dependent operators specifying dynamics of the coupled systems at the micro- and macro- levels, respectively.

The classical conveyor-tube model can be considered a particular case of the system-theoretic definition. There are coupled autonomous systems of two different types (i.e. *n*=2): psychic (i.e. humans – the "senders" and "receivers") and social (e.g. a language or medium/media). An observable state (i.e. realization) $R_2$ of the social system corresponds to a socially recognized (or anticipatedly effective) representation of a concept from $S_2$, a class of possible concepts. Observable states $R_1$ of the psychic systems represent (as "externalized," for instance, in the form of utterances) $S_1$ conceived perceptions and ideas, while the roles of the "senders" and "receivers" are explicitly established by ordering the corresponding time-intervals $[t_{j-1}, t_j]$. Figure 1 illustrates the coupling process: horizontal arrows describe the changing of the system states, while arrows crossing the system "borders" signify the perturbation of a system.

The social system works to filter, or authorize, communication out of human behavior and also to buffer the behavior against the uniformity of socio-cultural norms. The social system, however, does not (and cannot) impose a "standard" of communicative behavior. Instead, it serves to propagate among psychic (autonomous) systems regularities enabling coordination of their behavior that results in experiential "classification" of the systems' shared environment (i.e. the systems of the other type) into a set of attractor basins – recurring behavior clusters (or behavioral patterns). Apparently, the better the coordination of the psychic systems, the higher the efficiency of communication [17, 10]. (More concrete examples portraying communication as coordinated behavior can be found in reference [10].)

The generalized model (1) thus overcomes the principal flaws of the conveyor-tube framework: multiple media/channels, as well as senders and receivers can uniformly be described through their inner and observed states, while important characteristics of the discourse, such as (non)linearity, openness, and efficiency, can be controlled by changing properties of the state-realization selectors $\mathbf{E}$ and $\mathbf{I}$. It should also be stressed that the system-theoretic framework provides additional analytic power by shifting the modeling focus from the isolated coding and transmission processes to the overall dynamics of communication revealed at different (micro- and macro-) levels and in different domains (e.g. physical, cognitive, and social, when $n$=3). The latter allows for a wider range of meaningful scenarios for modeling the communication process.

## 3. Mathematical Interpretation

There are many perspectives from which the communication process can be modeled. In Shannon-Weaver mathematical theory of communication, the discussion is built around a single act of coding, transferring, and decoding information [21, 14]. To model the communication process, the knowledge of the channel/medium properties or coding/decoding procedures is then required but is usually beyond our reach and control in the case of hypermedia-based communication. Alternatively, in linguistic theories, the discussion is built around set-theoretic properties of representations, which constitute the discourse, and rules of the development of these representations that however seldom results in a traceable model (e.g. [2]). In the present study, we take yet a different perspective by focusing on the dynamics of observed realizations of coupled system macrostates.

We begin by postulating that the same macrostate (uniquely characterized by macroparameters) of a system involved into communication can have different realizations (just as the same perception or idea can be expressed in many ways, and the same representation can signify multiple concepts). We will also assume that while the number of macrostates is generally infinite (emphasizing the diversity of communication situations), the number of macrostate realizations (e.g. representations, such as words or hyperlink anchor texts) must be finite for at least one type of the coupled systems (1) to guarantee that the behavioral coordination revealed as, for instance, a correlation between naturally consecutive (i.e. as appeared) realizations in turn-taking communication can be achieved in a finite time. (This can be illustrated by the fact that traditional communication media, such as languages, utilize a finite number of socially recognized representation constructs – signs.) The latter assumption makes the communication process representationally cyclic: given a sufficient time, recurrence of identical realizations is inevitable.

We will proceed by calculating the distribution of $z$ the occurrence number of a state realization across increasing expenditures of time (e.g. it can be understood as the distribution of the occurrence frequency of a word in a document). Under the above assumptions, $z(t)$ can be defined as

$$z(t) = \theta \tau_0(t),$$ (2)

where $\tau_0(t)$ is the *realization rate*, and $\theta$ is the observation time – a behavioral macroparameter that may be associated with interaction tempo.

Owing to the local character of representing macrostates, there are multiple realizations of a given state. A number of statistically independent factors, including those due to semantics (i.e. idiosyncratic experience) and syntax (i.e. social norms), determine their rates $\tau_i(t)$, and some of these factors affect the observed realization by increasing or decreasing its rate $\tau_0(t)$. Temporal changes of $\tau_i(t)$ are controlled by a "competition" process, which can often be thought of as representational decision-making (conscious or otherwise), and in which different realizations compete for the time available to represent the macrostate. Taking into account the fact that macrostate realizations are usually uniquely allocated in time, changes of the available time – i.e. representation time dissipation – can be estimated by calculating the difference in the (macro)state and its realization rates. The dynamics of $\tau_i(t)$ can then be approximated with a diffusion process represented by stochastic differential equations (see reference [7] for a thorough introduction to diffusion processes; in reference [6], the authors discussed a similar diffusion mechanism):

$$\frac{d\tau_i}{dt} = a_i\left( \rho\mu - \sum_{j=0}^{N} \tau_j \right) + \eta_i,$$   (3)

where $\mu$ is the *macrostate rate*, $N$ is the number of competing realizations, $i=0,1,\ldots,N$; $\eta_i(t)$ is due to a noise-induced variation in realization rates and is a Gaussian stochastic variable with zero mean; $a_i(t)>0$, and $\rho \geq 1$ is a parallelism (or, in other words, redundancy) coefficient – a medium macroparameter to account for the apparent concurrency in representing the macrostate.

The system of differential equations (3) describes a diffusion process in the vicinity of $\rho\mu = \sum_{j=0}^{N} \tau_j$ the hyperplane that yields the uniform probability distribution with a probability density function (PDF) $\phi(x) = \dfrac{1}{\rho\mu}$ for $N$ independent realization rates $\tau_i(t)$ on this hyperplane. Taking into account equation (2) and for $N >> 1$, $F_1(z)$ a cumulative distribution function (CDF) of $z$ can be estimated through the marginal distributions of $z_i$, $i=1,\ldots,N$, as follows [6]:

$$F_1(z) \propto 1 - \left( 1 - \frac{z}{\theta\rho\mu} \right)^{N} \approx 1 - e^{-\lambda z},$$   (4)

where

$$\lambda = \frac{\varepsilon}{\theta\rho}, \qquad (5)$$

and $\varepsilon = \frac{1}{\tau}$ is the average realization time for $\tau = \frac{\mu}{N+1}$ the average realization rate of a given macrostate.

It is understood, that despite the obviously simplified character of the description of the "diffusion" of a macrostate over its observed realizations with equations (3), the exponential behavior of equation (4) holds for a wide class of models, e.g. when some of the parameters change with time, when some of the factors determining realization rates depend on each other, etc.

By extending the obtained result to the case of a (large) number of different macrostates simultaneously observed throughout the communication process, it is easy to derive:

$$f_0(z) = \int_0^\infty \varphi(\lambda)\lambda\,\mathrm{e}^{-\lambda z}\,d\lambda, \qquad (6)$$

where $f_0(z)$ is the PDF of $z$, and $\varphi(\lambda)$ is the PDF of $\lambda$.

The latter generalization partially accounts for the fact that in the case of complex (autonomous) systems, a measured stochastic variable ($z$, in our case) reflecting a system's behavior is, as a rule, a sum of random variables, where each of the summands stands for the system's behavior in a steady or stationary state with certain parameters of the system's internal regulatory mechanisms. $f_0(z)$ thus specifies the distribution function of the realization occurrence number for the states characterized by the existence of different macroparameters shaping a single distribution of $\lambda$. However, since communication encompasses coupled systems of different nature (e.g. cognitive, social, and physical) and with different properties that may well be expected to produce different distributions of $\lambda$, the PDF of $z$ should be reformulated to a still more universal form:

$$f(z) = \sum_{i=1}^{M} c_i f_{0_i}(z),\qquad (7)$$

where $M$ is the number of distributions of $\lambda$ (it can often be presumed that $M$ is equal to $n$ the depth of coupling; usually, $M$=2 or 3), and $c_i$ gives the probability to observe the "$i$-th type system" states in communication.

So far, we did not make any assumption about $\varphi(\lambda)$ the PDF of the composite parameter $\lambda$, and our result – the distribution equation (7) – is fairly general but is difficult to apply in practice because it yields no specific functional form. It is clear from equation (5) that as long as $\theta$ and $\rho$ are constant (or change relatively slowly), the distribution of $\lambda$ depends on that of the realization time $\varepsilon$. The latter is a stochastic variable characterizing, at least in the case of human communication, the durations of higher nervous activities, the exact distribution function of which is not known. There is substantial empirical evidence, however, that a Gamma distribution provides a reasonably good approximation [13]. By substituting the Gamma PDF

$$\varphi(\lambda) = \frac{b^{\nu}}{\Gamma(\nu)} \lambda^{\nu-1} e^{-b\lambda}, \quad \lambda \ge 0, b > 0, \nu > 0, \Gamma(\cdot) \text{ denotes the Gamma function, into}$$

equation (6) and after integration, equation (7) can be re-written as

$$f(z) = \sum_{i=1}^{M} c_i \frac{v_i \, b_i^{v_i}}{(z + b_i)^{v_i+1}} \, , \qquad (8)$$

where $b_i$ and $v_i$ are distribution parameters. In terms of equation (4), as the expectation

(i.e. mean) $\mathrm{E}[\lambda] = \dfrac{v}{b}$ for $\lambda$ a Gamma distributed variable, $\dfrac{v_i}{b_i} \propto \dfrac{1}{\theta_i \rho_i} \mathrm{E}[\varepsilon_i]$, $i = 1, \ldots, M$.

Thus, it finally turns out that the probability density function of $z$ the state realization

occurrence number in communication is specified by the finite sum of Pareto Second

Kind (Lomax) PDFs (8) with parameters, which may be associated with higher nervous activities ($\varepsilon$), behavioral characteristics ($\theta$), and efficiency of representation ($\rho$). In the

following, we will show that this result is not only empirically superior to the one

obtained within the conveyor-tube classical framework, but it also provides us with a

new insight into the analysis of hypermedia systems and communication. (It is

interesting to note that the Pareto 1 distribution can be obtained from the general

formula (8) as a special case when $M = 1$ and $z \gg b$.)

## 4. Comparison with Empirical Data

To compare the reported theoretical findings with natural text-based and hypermedia-

based communication data, we have collected a selection of texts and hypertexts. The

texts are Associated Press articles on politics retrieved through and downloaded from

`http://www.washingtonpost.com/wp-srv/searches/mainsrch.htm.`

Four selections of hypertexts on politics were downloaded from four different sites of a subjectively different level of multimedia deployment:

`http://www.cbsnews.com/sections/politics/main.shtml` (CBS News, "moderate-to-high" multimedia),

`http://www.cnn.com/ALLPOLITICS/` (CNN, "high" multimedia),

`http://dmoz.org/Society/Politics/News_and_Media/` (Open Directory Project – DMOZ, "low" multimedia), and

`http://dir.yahoo.com/Government/Politics/` (Yahoo, "low" multimedia).

Nodes comprising the hypertexts were connected with each other by at least one hyperlink within the site.

All the article texts were raw English texts, where we ignored the punctuation and turned all word-forms into lower-case; different word-forms were left unchanged and thus treated as different realizations of a macrostate. All the hypertexts were stripped of multimedia and technical contents, so that only English texts remained, which were then processed in the same way as the Associated Press articles. The obtained five collections were of a size of approximately 10 Megabytes each.

We have calculated the occurrence number of words in the processed texts and hypertexts and thus created five data samples. These latter samples represent results of the "measurement" of the occurrence frequency of a word made at different times in four hypermedia-based and one text-based communications. The sample size varied from approximately 1 million words (CNN, Open Directory, and Yahoo) to 1.8 million (CBS), and to 2 million words (Associated Press).

A preliminary analysis has shown that all the data sets exhibit extremely skewed behavior: very few occurrence frequencies dominate the samples so that words met only once or twice constitute up to 60% of the words used in communication. At the same time, high word frequencies have a very low probability but still cannot be neglected as they correspond to most popular (i.e. most used) words. All this means that statistical estimates, such as expectation (mean) and variance, become virtually useless, but a distribution function must be found to characterize this data. The latter task – finding a PDF fit to the data – is of utmost importance in information retrieval, data mining, and information system design (see reference [20]; reference [12] – for a theoretical justification). Zipf's "law" and its mathematical interpretation (by Mandelbrot [14]) in the form of the Pareto 1 distribution are habitually applied as a de facto standard for modeling such data in computer science. It then appears natural to test both the classical (Pareto 1 distribution) and the proposed (sum of Lomax distributions) analytical models against the data.

Maximum likelihood estimators were used to derive parameters of the models (the necessary technical adjustments have been made in the models to deal with discrete data). In the case of the sum (8), the coefficients $c_i$ as well as the number of summands $M$ were sought through a pareto-optimization procedure by minimizing $\chi^2$ the chi-square statistic. $\chi^2$ was also used to test the goodness-of-fit as the only legitimate statistic to be applied to discrete distributions. To ensure the validity of the chi-square approximation, bins naturally formed by the discrete data in the right tail of the distributions, which contained less than 6 elements, have been merged.

In our first experiment, we dealt with the entire data samples. It has been found, that the Pareto 1 distribution does not fit the empirical data at any acceptable level of significance. On the other hand, the proposed system-theoretic model provides a very good fit to the data in all five cases with a significance level α varying from 0.05 (the CBS hypertexts; the best fit) to 0.01 (natural text, CNN, Yahoo, and Open Directory). Figures 2 and 3 present examples of the modeling with insets displaying the results on a log-log scale (in Figure 3 – in comparison with the classical model).

Table 1 lists the proposed model parameters along with $\frac{v_i}{b_i} \propto \mathrm{E}[\lambda_i]$ calculated in the experiment.

For our second experiment, we have prepared 30 data samples by dividing the CBS, natural text, and DMOZ collections into 10 subsets each. The size of the subsets varied from 0.5 to 1.0 million words. Using this data, we have calculated estimates of the parameters of the model (8).

It has been found that $M$=2 provides a reasonably good fit in all 30 trials. The $c_i$ coefficients fluctuated around $\overline{c}_1 = .56$ and $\overline{c}_2 = .44\,(\pm.08)$. No significant correlation has been detected between sample size and any of the model parameters. Figure 4 depicts the scatter plot of the mean values $\mathrm{E}[\lambda_i]$ for the 30 samples. In terms of $b_i$ and $v_i$, there is a significant difference ($P$<0.001) between the CBS and DMOZ collections, while the natural text is positioned "in between" and overlaps with these two.

## 5. Discussion

Figures 2 and 3 demonstrate that the developed analytical model fits the word occurrence distribution very well in the case of both word- and hypermedia- based communication. The fault of the classical model is, on the other hand, not surprising: while even intuition suggests that characterization of communication in terms of stationary statistics of encoded messages is not general enough to cover the vast diversity of communication means (e.g. hypermedia), building the discussion around the optimization of encoding procedures appears poorly motivated if not wrong at all. Accordingly, although the Zipf-Mandelbrot prediction of the Pareto 1 distribution for word frequency (i.e. notorious Zipf's second "law") has become (perhaps, owing to its transparency) a classical model in many domains, not limited to linguistics, we claim that often this power law is a popular and convenient "belief" rather than a model that can be validated with statistical rigor. In spite of the abundant literature on the subject (see reference [8, 16] for bibliography), there is little evidence that Zipf's law holds for any real communication data: very few works provided statistical validation of the "law" in its original form, which however was a result of more or less sophisticated manipulations, such as "noise" filtration, rather than inherent property of the data samples.

Expectedly, many refinements have been proposed over the past decades that help improve the classical model to better conform with empirical data. The main problem of these improved models is that they usually remain empirical by its very nature and, therefore, even when they do provide very good fits (e.g. as in reference [5]), their parameters can hardly be interpreted in terms of the communication process. The empirical modes fall short of convincingly explaining most of the communication properties that they discover, for instance – the existence of two distinct domains of word frequencies [16], and thus end up as *ad hoc* and inconsistent.

As an alternative to the classical information (as encoded messages) transmission framework, the system-theoretic modeling provides not only a more general perspective on the communication process and a better fit to the empirical data, but it also offers a meaningful interpretation of the model parameters that can be used in the design of information systems. In particular, it is now understood that distinct domains of word frequencies can be due to the differences in the parameters of the coupled systems, and their number is not generally limited to 2. (As one can see from the inset in Figure 3, the model fit could perhaps be improved by adding one more summand, e.g. in order to account for the coupling with the physical environment, to the distribution equation. This addition does not, however, seem necessary as even in its simplest form, i.e. with two summands, the model is statistically sound.)

It is important to note that the values of the $c_i$ coefficients obtained in our experiments are in a good agreement with the fact that (natural) language words can generally be classified into "service" and "content" words, which make up 40-45% and 55-60% of the (English) lexicon, respectively. We could then speculate that service words are products of the social system, whereas content words are realizations (representations) of cognitive system macrostates. Another interesting finding is that in terms of word usage, communication does not significantly change its properties with time: from certain lengths (sizes) up, there is no stable correlation between length of communication (i.e. sample size) and values of the model parameters. On the other hand, these parameters have appeared to be good indicators of medium dynamics: based on the second experiment results, we could speculate that in hypermedia-based communication, multimedia-induced (redundant) content helps maintain the dynamics similar to the one occurring in natural text. We can also expect that statistical characteristics of anchor text are different from the ones of natural text. The latter may affect the effectiveness of the algorithms used by search engines, data mining tools, and other applications sustaining hypermedia-based communication.

These two findings – the detection of the two domains in the English lexicon and the discerning between the dynamics of word- and hypermedia- based communication – are critical for validation of our analytic framework. Indeed, the proposed model (8) is fairly complex, and it would simply overfit the empirical data (overfitting happens when the flexibility of a model allows it to fit data to such high accuracy that the fit is driven by the random fluctuations in the data). While the large number of parameters in the system-theoretic model would compromise its predictive capability, the obtained experimental results permit us to claim that the model does describe the communication process, rather than merely approximating the data. It is understood, however, that the application of the model to analyze data of a different nature (e.g. non-English texts or web page hits) may require finding additional supportive arguments.

The work in this paper was motivated in part by our inability to apply the results previously obtained by researchers working on various aspects of statistical description of communication to the analysis of the evolution of representation media (e.g. the evolution from interpersonal to mass-media, and towards hypermedia-based and ubiquitous computing communication). One problem is that the known analytical models exclusively deal with letters and words and, strictly speaking, cannot and should not be extended to the case of other representations [14]. Another problem is that most evolutionary models, such as the Price fundamental equation [19], can hardly utilize the Zipf-Mandelbrot "law," as parameters of the latter are thought to reflect some static characteristics of the discourse (e.g. the openness/closedness of the vocabulary) but have no clear evolutionary meaning. Besides, moments of probability distributions, which usually play a central role in evolutionary models, cannot often be evaluated analytically in the case of power-law distributions because the corresponding integrals diverge. The proposed system-theoretic approach does not place any restriction on the nature of representations (i.e. macrostate realizations), as it approaches communication from a very general behavioristic point of view. It therefore predicts that fluctuations of representation frequencies will follow the law specified by equation (8) not only in letter-by-letter/word-by-word, but in any other "representation-based" communication.

The evolution of representation media can simplistically be described with the Price reduced equation [19, 9] as $\Delta \bar{z} = z_{fr} - \bar{z} + \Delta z_{fr}$, where $\Delta \bar{z}$ is the rate of cultural evolution of the medium $z$ – the medium (e.g. language) is likely to eventually become extinct if $\Delta \bar{z} < 0$ for a majority of its elements (e.g. words); $z_{fr}$ is a fresh-hold specifying the usage frequency that "guarantees" the selection and reuse of a representation (e.g. a keyword frequency for a document), $\bar{z}$ is the average representation frequency in communication, and $\Delta z_{fr}$ is the selection error (e.g. due to differences in indexing algorithms). The mean of a Lomax-distributed random variable is, in terms of equation (8), equal to $\frac{b}{\nu}, \nu > 0$. Taking into account the fact that only $\bar{z}$ is the constantly negative term in the Price reduced equation, we could speculate, based on the results of our second experiment (see Figure 4), that under other similar conditions, hypermedia representations (e.g. links with anchor text) are less likely to extinct at sites devoid of multimedia redundant representations. On the other hand, the presence of a rich multimedia content may diversify the words used on a particular site over time. While a detailed analysis of the evolution of communication is beyond the scope of our paper, this oversimplified example demonstrates the potential of the system-theoretic approach for the analysis and design of information systems.

## 6. Conclusions and Future Work

The work reported in this paper reflects a particular view of the communication process, which is currently not the mainstream in computer science and linguistics. We formulated a system-theoretic framework for modeling communication, developed its mathematical interpretation, and applied the obtained model to predict and analyze some of the important characteristics of word- as well as hypermedia- based communication and information systems. To summarize our main findings, we have argued that 1) in the case of hypermedia systems, the communication process and, specifically, the representation dynamics cannot be modeled using a single power-law distribution, and the Zipf-Mandelbrot "law" does not hold; 2) even in the most general case, the dynamics of communication as manifested by the deployed representation frequencies can be modeled in terms of a sum of Lomax (also called Pareto Second Kind) distributions. The reported theoretical results have also made explicit the dependence between social and cognitive parameters of communication.

Given the popularity of the Zipf-Mandelbrot approximation and its analogs developed in many domains, such as linguistics, economics, and biology (e.g. see reference [3, 22]), it may be expected that some of the power (Zipfian) "laws" be fundamentally due to various "diffusion" processes caused by (conscious or otherwise) choice and decision-making, just as it was shown in the present paper. Our future work consists of extending and validating the developed mathematical model for other seemingly "scale-free" phenomena (e.g. as in communication networks [1]), and also applying the results reported herein to study the evolution of representation media.

## 7. Acknowledgement

Two of the authors (K.V.V. and K.K.) are receiving support from the Japan Science and Technology Agency, "The Universal Design of the Digital City" project.

## References

1. Adamic L., Network Dynamics: The World Wide Web, Ph.D. Thesis, Stanford University, June, 2001,

`http://www.webir.org/resources/phd/Adamic_2001_thesis.pdf.`

2. Andersen P.B., Dynamic Semiotics, *Semiotica* 139 (2002) 161-210.

3. Barabasi A., Albert R., Emergence of scaling in random networks, *Science Magazine* 286 (1999) 509.

4. Bardini T., Bridging the Gulfs: From Hypertext to Cyberspace, *Online Journal of Computer-Mediated Communication, JCMC*, 3(2) (1997)

`http://www.ascusc.org/jcmc/vol3/issue2/bardini.html.`

5. Bi Z., Faloutsos C., Korn F., The DGX distribution for mining massive, skewed data, in: *Proceedings of the Seventh ACM SIGKDD International Conference on Knowledge Discovery and Data Mining (KDD),* S.-F. Calif., USA, 2001, pp. 17-26.

6. Czirok A., Schlett K., Madarasz E., Vicsek T., Exponential distribution of locomotion activity in cell cultures, *Physical Review Letters*, 8 (14) (1998) 3038-3041.

7. Ghez R., *Diffusion Phenomena. Cases and Studies*, Kluwer Academic Publishers, Dordrecht/ Boston/ London, 2001.

8. *Glottometrics*, Special issue to honor G.K. Zipf, 4 (2002) `http://www.ram-verlag.de.`

9. Henrich J., Boyd R., On Modeling Cognition and Culture, *Journal of Cognition and Culture*, 2 (2) (2002) 87-112.

10. Kryssanov V.V., Okabe M., Kakusho K., Minoh M., A theory of communication for user interface design, in: Henk W.M. Gazendam, René J. Jorna, and Ruben S. Cijsouw (eds), *Dynamics and change in organizations: studies in Organizational Semiotics*, Kluwer Academic Publishers, Dordrecht/ Boston/ London, 2003, pp. 37 –59.

11. van Leijenhorst D.C., van der Weide Th.P., A formal derivation of Heaps Law, *International Journal of Information Sciences*, (2004, in press).

12. Losee R.M., Term Dependence: A Basis for Luhn and Zipf Models, *Journal of the American Society for Information Science and Technology*, 52 (12) (2001) 1019-1025.

13. Luce R.D., *Response Times. Their Role in Inferring Elementary Mental Organization*, Oxford University Press, New York, 1986.

14. Mandelbrot B., On the Theory of Word Frequencies and on Related Markovian Models of Discourse, in: Roman Jakobson (Ed.), *Proceedings of the Twelfth Symposium in Applied Mathematics*, American Mathematical Society, New York, 1960, pp. 190-219.

15. Maturana H., Varela F.J., *Autopoiesis and Cognition: The Realization of the Living*, D. Reidel Publishing Company, Dordrecht, 1980.

16. Naranan S., Balasubrahmanyan V.K., Information theoretic models in statistical linguistics – Parts I and II, *Current Science*, 63 (1992) 261-269 and 297-306.

17. di Paolo E.A., An investigation into the evolution of communication, *Adaptive Behavior*, 6(2) (1998) 285-324.

18. Pitkow J.E., Summary of WWW Characterizations, *Web Journal* 2(1-2) (1999) 3-13.

19. Price G.R., Selection and Covariance, *Nature*, 227 (1970) 520-521.

20. van Rijsbergen C.J., *Information Retrieval*, Butterworths, London, 1979.

21. Shannon C. E., Weaver W., *The Mathematical Theory of Communication*, University of Illinois Press, Urbana, Illinois, 1949.

22. Shide N., Batty M., Power Law Distributions in Real and Virtual Worlds, in: *Inet 2000 Proceedings*, Internet Society, 2000, `http://www.isoc.org/inet2000.`

23. Simon H.A., On a class of skew distribution functions, *Biometrika*, 42 (1955) 425-440.

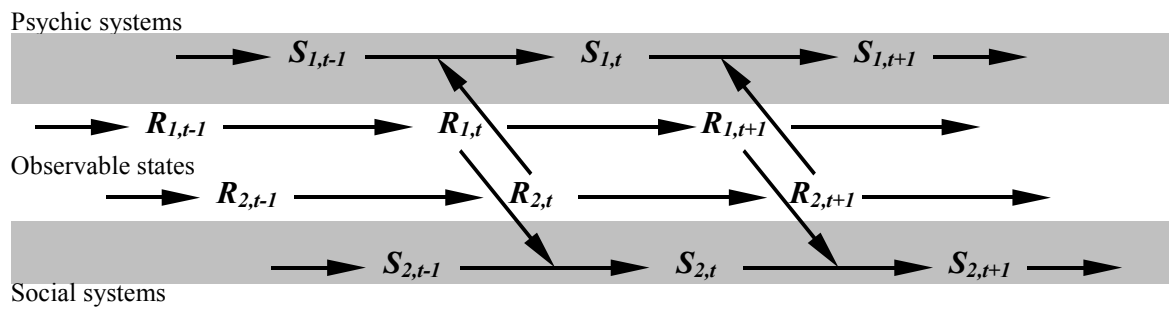

Figure 1: Communication as coupling of autonomous systems (the case when the depth $n = 2$).

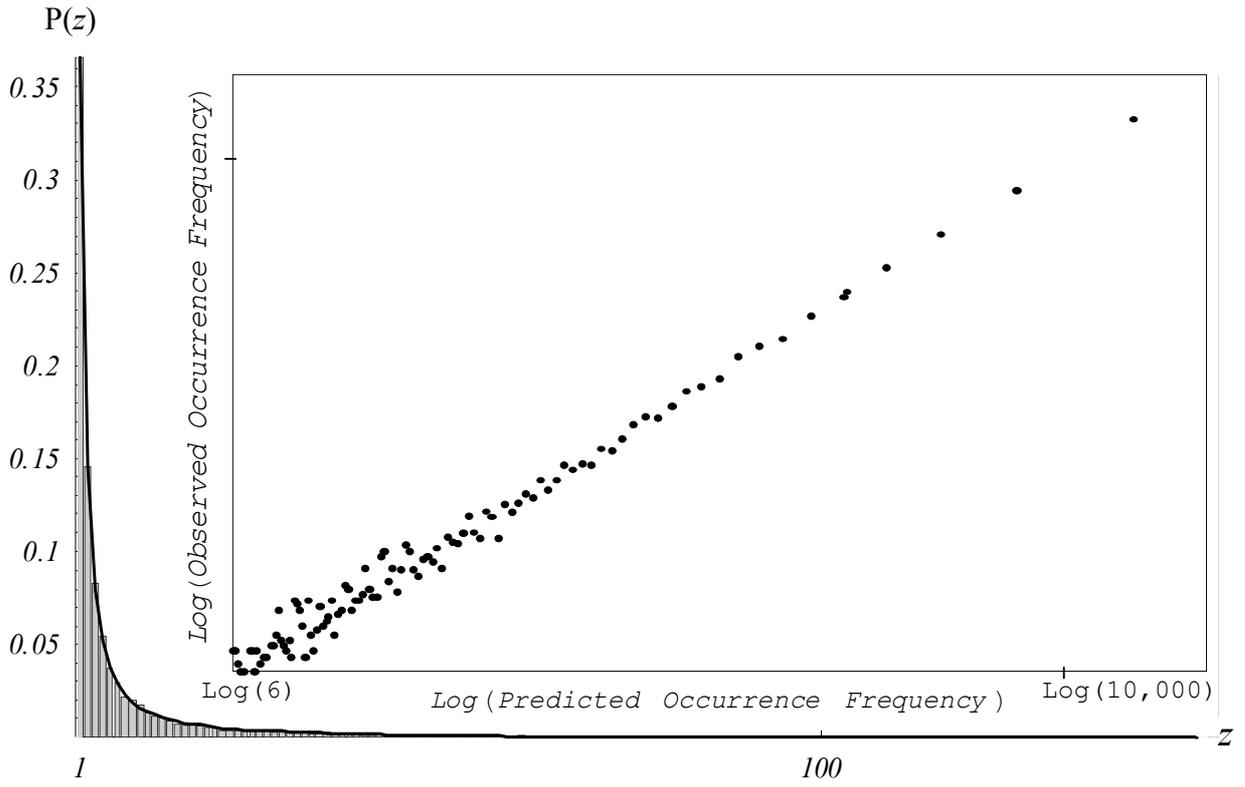

Figure 2: The frequency of word occurrence in the CBS sample (histogram) and its prediction (solid curve) by the system-theoretic model with parameters estimated from the data (see Table 1). In the inset: histogram of the predicted occurrence frequency versus the observed occurrence frequency plotted on a log-log scale.

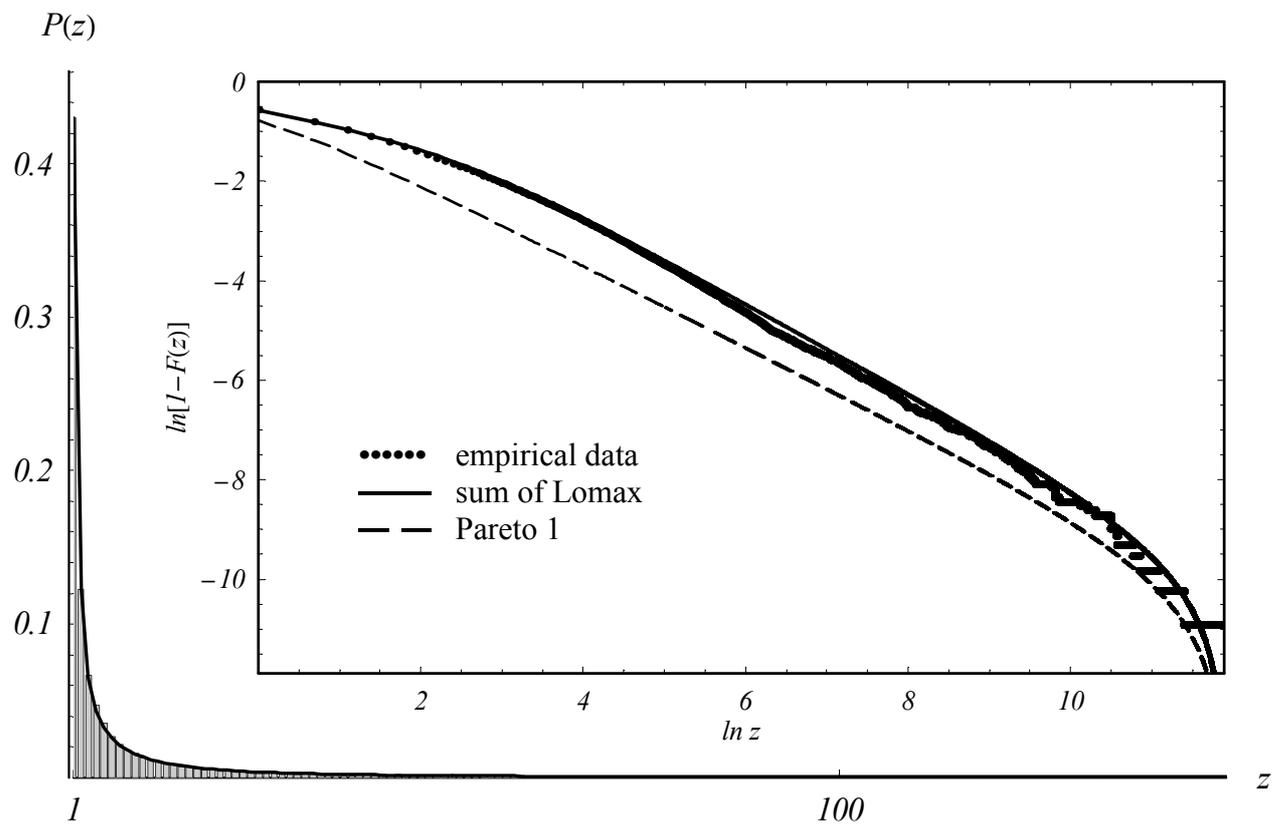

Figure 3: The frequency of word occurrence in the "natural text" sample (histogram) and its prediction (solid curve) by the system-theoretic model. In the inset: the distribution functions plotted for the full data range and on a log-log scale; CDF

$$F(x) = P(X \le x) = \int\limits_0^x f(t)dt.$$

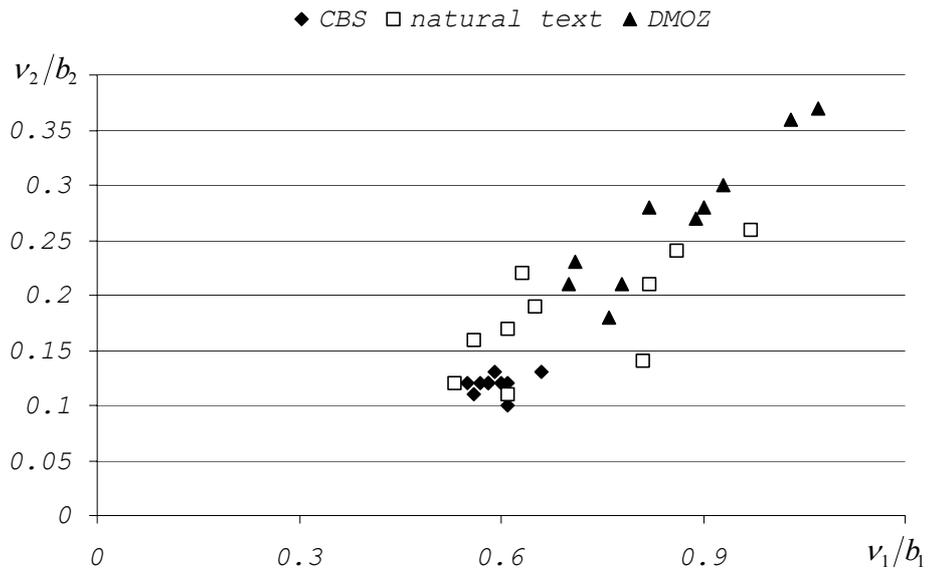

Figure 4: Grouping of the model parameters for the different samples.

| Sample | M | $c_1$ | $c_2$ | $v_1$ | $b_1$ | $v_2$ | $b_2$ | $\dfrac{v_1}{b_1}$ | $\dfrac{v_2}{b_2}$ |
|---|---|---|---|---|---|---|---|---|---|
| CBS | 2 | 0.55 | 0.45 | 1.19 | 2.08 | 0.89 | 7.26 | 0.57 | 0.12 |
| CNN | 2 | 0.62 | 0.38 | 1.01 | 1.78 | 0.96 | 9.51 | 0.57 | 0.10 |
| Natural text | 2 | 0.53 | 0.47 | 1.83 | 1.02 | 0.90 | 6.63 | 0.81 | 0.14 |
| Yahoo | 2 | 0.49 | 0.51 | 7.52 | 4.50 | 1.41 | 3.73 | 1.67 | 0.38 |
| DMOZ | 2 | 0.78 | 0.22 | 0.46 | 0.65 | 0.90 | 4.27 | 0.70 | 0.21 |

Table 1: Fit model parameters.